\documentclass[a4paper,10pt]{article}
\usepackage{amsfonts}
\usepackage{graphicx}
\usepackage{amsmath}
\usepackage{amssymb}
\usepackage[breaklinks=true,dvips]{hyperref}
\usepackage{breakurl}
\usepackage[utf8x]{inputenc}
\usepackage{mathrsfs}
\usepackage{color}

\title{\bf Non-orthogonality and $\kappa$-dependence eccentricity of polarized electromagnetic waves in CPT-even Lorentz violation}
\author{Thiago Prud\^encio\footnote{e-mails: thprudencio@gmail.com; prudencio.thiago@ufma.br} \\ 
\small
\it Coordination of Science \& Technology (CCCT/BICT)
\\ 
\it \small  Federal University of Maranh\~ao (UFMA), \\ 
\small \it 
65080-805, Sao Lu\'is-MA, Brazil. \\
\normalsize \\
Humberto Belich\footnote{e-mail:belichjr@gmail.com}\\
\it \small Departamento de F\'isica e Qu\'imica (DFQ), \\
\it \small Federal University of Espirito Santo (UFES), \\
\it \small 29060-900, Vitoria-ES, Brazil.}

\date{}

\begin{document}

\maketitle

\begin{abstract}
We discuss the modified Maxwell action of a $K_{F}$-type Lorentz symmetry breaking theory and present a solution of Maxwell equations derived in the cases of linear and elliptically polarized
electromagnetic waves in the vacuum of CPT-even Lorentz violation. We show in this case  
the Lorentz violation has the effect
of changing the amplitude of one component of the magnetic field, while leaving
the electric field unchanged, leading to non-orthogonal propagation of eletromagnetic fields and dependence of the eccentricity on 
$\kappa$-term. Further, we exhibit numerically the consequences of this effect in the cases of linear and elliptical
polarization, in particular, the regimes of non-orthogonality of the electromagnetic wave fields and the eccentricity of the elliptical polarization of the
magnetic field with dependence on the $\kappa$-term. 
\end{abstract}
{\small
\tableofcontents
}

\section{Introduction}

The recent discovery of the Higgs boson at the LHC established a research
program that aims to explain the physics of fundamental interactions, like
excitations manifested from fundamental fields, ending a program of
tremendous experimental predictions - the Standard Model (SM) of particle
physics. This model has the merit of unifying the weak force and
electromagnetism by the Anderson-Higgs mechanism. Nevertheless, it is known
that the description of the SM without neutrino mass is unsatisfactory.
Recent experiments indicate that the electrons have a bound for the electric
dipole moment \cite{dip,baron,bernreuther} and the SM predicts that the electron is a punctual
particle. By these experimental results we understand the necessity of
investigating physics beyond the Standard Model. Kostelecky and Samuel \cite{sam} showed that 
interactions in the context of
the string field theory could lead to spontaneous breaking
of the Lorentz symmetry.  Such a scenario cannot
exist in the regime of SM, but can be the key for
explaining regimes where quantum gravity and the early stages of the
universe bring remanecent breaking effects in the limits of validity of SM.
In these other regimes, structure of spacetime and the corresponding particle
physics are affected in their symmetries. A spontaneous breaking of Lorentz symmetry can then be extented beyond the known
scenarios of QED and QCD \cite{frohlich,balachandran0,balachandran1,balachandran2,jenkins}. 
 
Lorentz violation theories are in the framework of Standard
Model Extension (SME) \cite{kostelecky0,kostelecky1}. 
From the SM perspective, Lorentz symmetry breaking induces an assymmetry in
the structure of spacetime, implying for instance that vacuum Maxwell
equations has to be changed by non-trivial contributions. One important feature of LVT in
the domain of electrodynamics is the possibility of violating Lorentz
symmetry without breaking gauge symmetry. One of the first models in
this sense was carried out by the Carroll-Field-Jackiw model \cite{carroll} where breaking
of the Lorentz invariance is realized by the inclusion of an additional term
in the Maxwell lagrangian, while preserving the gauge symmetry. General case of $\kappa$-electrodynamics was introduced in \cite{colladay}, determining 
the energy momentum tensor and
basic properties of the plane wave solutions of $\kappa$-electrodynamics. A more in-depth study of
the model and its plane wave solutions was performed in \cite{mewes}, where equations of motion in Maxwell
form are derived. A special case of the $\kappa$-parameter was discussed in \cite{schreck} that also constructs the plane-wave solutions
including with polarization vectors. The idea that the special case of
the $\kappa$-parameter acts like an effective metric has been used widely
in the literature \cite{cambiasso}. 

The presence of terms that violate the Lorentz symmetry imposes at
least one privileged direction in the spacetime \cite{geral,geral1,geral2,geral3,geral4}. Nowadays,
studies in relativistic quantum effects \cite{knut1,knut11,knut12,knut13,knut14} that stem from a non
minimal coupling with Lorentz symmetry breaking \cite{knut21,knut22,knut23,knut24,knut25} has opened the
possibility of investigating new implications in quantum mechanics that this
violator background can promote \cite{ens1,ens2,ens3}. 

In this work, we discuss free electromagnetic fields with a
particular type of Lorentz symmetry violation: the vacuum reduction of a $K_{F}$-type Lorentz
symmetry breaking theory \cite{belich} in the absence of complex scalar
field $\varphi $ and $\lambda $-parameter associated to a $\lambda |\phi
^{4}|$ are zero. The modified dispersion
laws of electromagnetic waves are obtained in \cite{4}
with two four-vectors as violating background, i.
e., they are interpreted as preferred directions in
spacetime. We investigate the scenario of the the modified Maxwell
equations resulting from this CPT-even scenario of Lorentz violation. In
this case, we consider the modified Maxwell equations in the vacuum of a $K_{F}$-type Lorentz symmetry breaking where the Lorentz violation is manifested only by the presence of a $\kappa ^{\mu \nu \rho \sigma }$ tensor. We then consider a solution of Maxwell equations derived in the cases of linear and elliptical polarization
electromagnetic waves in the vacuum of CPT-even Lorentz violation scenario and investigate the effect
of the amplitude changing of the amplitudes electromagnetic wave fields and regimes of non-orthogonality of the electromagnetic wave fields and the eccentricity of the elliptical polarization with dependence on the $\kappa$-term.  

The structure of this paper is the following: in section II we discuss the modified Maxwell action by the presence of the tensor {$\kappa_{abcd}$}; in section III, the discuss the lagrangian; in section IV, the effective metric; in section V, the Momentum Energy Tensor for the modified Maxwell action and in the section VI, the
Maxwell equations from the modified Maxwell lagrangian. In section VII, we present our
conclusions.

\section{The tensor {\Large $\kappa_{abcd}$}}

We starting off our analysis with the Maxwell action in the presence of the Lorentz violating term 
\begin{equation}
\Sigma =\int d^{4}x\left\{ -\sqrt{-g}\frac{1}{4}F_{\mu \nu }F^{\mu \nu }\right\}
+\Sigma _{k},  \label{1}
\end{equation}%
where the part responsible by the Lorentz violation is given by
\begin{equation}
\Sigma _{k}=-\frac{1}{4}\int d^{4}x \sqrt{-g}\left( \kappa ^{\mu \nu \rho \sigma
}\,F_{\mu \nu }F_{\rho \sigma }\right).  \label{2}
\end{equation}%
The tensor $\kappa_{\mu\nu\sigma\rho}$ is CPT-even, that is, it does not violate the
CPT-symmetry. Although the violation of the CPT-symmetry implies that the
Lorentz invariance is violated \cite{greenberg}, the reverse is not
necessarily true. The action with the presence of $\kappa_{abcd}$ breaks the Lorentz
symmetry in the sense that the tensor $\kappa_{abcd}$ has a non-null vacuum
expectation value. Besides, the tensor $\kappa_{abcd}$ has the same properties of
the Riemann tensor, as well as an additional double-traceless condition.
This tensor possesses the following symmetries: 
\begin{equation}
\kappa_{abcd}=\kappa_{\left[ ab\right] \left[ cd\right] };\,\,\,\kappa_{abcd}=\kappa_{cdab};\,\,%
\,\kappa_{\,\,\,\,\,\,\,ab}^{ab}=0.  \label{1.2}
\end{equation}

By following Refs. \cite{curv3,curv2,curv4}, we can write the tensor $%
\kappa_{abcd}$ in terms of a traceless and symmetric matrix $\tilde{\kappa}_{ab}$
as 
\begin{equation}
\kappa_{abcd}=\frac{1}{2}\left(\eta_{ac}\,\tilde{\kappa}_{bd}-\eta _{ad}\,\tilde{%
\kappa}_{bc}+\eta_{bd}\,\tilde{\kappa}_{ac}-\eta_{bc}\tilde{\kappa}%
_{ad}\right).  \label{123kkr}
\end{equation}

By defining a normalized parameter four-vector $\xi ^{a}$, which satisfies
the conditions: $\xi _{a}\xi ^{a}=1$ for the timelike case and $\xi _{a}\xi
^{a}=-1$ for the spacelike case; thus, we can decompose the tensor $\tilde{%
\kappa}_{ab}$ as 
\begin{equation}
\tilde{\kappa}_{ab}=\kappa \left( \xi _{a}\xi _{b}-\frac{\eta _{ab}\,\xi
^{c}\xi _{c}}{4}\right) ,  \label{1.4}
\end{equation}%
where 
\begin{eqnarray}
\kappa =\frac{4}{3}\tilde{\kappa}^{ab}\,\xi _{a}\,\xi _{b}.
\end{eqnarray} 
We deal with a Lorentz symmetry violating tensor $k_{\mu \nu \kappa \lambda }
$ in such a way that the CPT symmetry is preserved. It has been shown in
Ref. \cite{belich}  that a particular decomposition \cite{curv4,curv3} of
the tensor $\kappa_{\mu \nu \rho \sigma }$ produces a modification on the
equations of motion of the electromagnetic waves due to the presence of
vacuum anisotropies, which gives rise to the modified Maxwell equations. As
a consequence, the anisotropy can be a source of the electric field and,
then, Gauss's law is modified. Besides, the Amp\`{e}re-Maxwell law is
modified by the presence of anisotropies and it has a particular interest in
the analysis of vortices solutions since it generates the dependence of the
vortex core size on the intensity of the anisotropy.

\section{Lagrangian}
We start with the following modified Maxwell lagrangian from the CPT-even $K_{F}$-type Lorentz symmetry breaking theory
\begin{eqnarray}
\mathcal{L}_{\text{modMax}}\ &=& -\sqrt{-g}\,\left( \frac{1}{4}\,F_{\mu \nu
}F_{\rho \sigma }\,g^{\mu \rho }g^{\nu \sigma }+\frac{1}{4}\,\kappa ^{\mu
\nu \rho \sigma }\,F_{\mu \nu }F_{\rho \sigma }\right) \ . \nonumber \\
\label{eq:LagMMgrav-L}
\end{eqnarray}
Using the description of $\kappa ^{\mu \nu \rho \sigma }$ in tetrad fields 
\begin{eqnarray}
\kappa ^{\mu \nu \rho \sigma } = \kappa ^{abcd}\,e_{\;\;a}^{\mu
}\,e_{\;\;b}^{\nu }\,e_{\;\;c}^{\rho }\,e_{\;\;d}^{\sigma }\,
\end{eqnarray}
and the decomposition in the non-birrefringent sector eq. (\ref{123kkr}),
we can write the modified contribution as
\begin{eqnarray}
\kappa ^{\mu \nu \rho \sigma }\,F_{\mu \nu }F_{\rho \sigma } &=&\frac{1}{2}( \eta ^{ac}\widetilde{\kappa }^{bd}-\eta ^{ad}\widetilde{\kappa}^{bc} 
+ \eta ^{bd}\widetilde{\kappa }^{ac} \nonumber\\ 
&-&\eta ^{bc}\widetilde{\kappa }
^{ad}) \,e_{\;\;a}^{\mu }\,e_{\;\;b}^{\nu }\,e_{\;\;c}^{\rho
}\,e_{\;\;d}^{\sigma }\text{ }F_{\mu \nu }F_{\rho \sigma }.
\end{eqnarray}
Taking into account the relation between metric and tetrad fields
\begin{eqnarray}
g^{\mu \rho }e_{\;\;b}^{\nu
}e_{\;\;d}^{\sigma }\text{ }&=& \eta ^{ac}e_{\;\;a}^{\mu }\,e_{\;\;b}^{\nu
}\,e_{\;\;c}^{\rho }\,e_{\;\;d}^{\sigma }\text{ },
\end{eqnarray}
we also have
\begin{eqnarray}
\kappa ^{\mu \nu \rho \sigma }\,F_{\mu \nu }F_{\rho \sigma } &=& \frac{1}{2}( g^{\mu \rho }\widetilde{\kappa }^{bd}e_{\;\;b}^{\nu
}e_{\;\;d}^{\sigma }\text{ }-g^{\mu \sigma }\widetilde{\kappa }%
^{bc}e_{\;\;b}^{\nu }\,e_{\;\;c}^{\rho }\,\text{ } \nonumber \\
&+& g^{\mu \nu }\widetilde{\kappa }^{ac}\,e_{\;\;c}^{\rho }\,e_{\;\;d}^{\sigma }\text{ }\nonumber \\
&-& g^{\nu \rho }%
\widetilde{\kappa }^{ad}e_{\;\;a}^{\mu }\,e_{\;\;d}^{\sigma }\text{ })\,F_{\mu \nu }F_{\rho \sigma }. 
\end{eqnarray}
Considering the tensor 
\begin{eqnarray}
\tilde{\kappa}^{ab}=\kappa \left(\xi^{a}\xi^{b}-\eta^{ab}\xi^{c}\xi_{c}/4\right),
\end{eqnarray}
we can rewrite 
\begin{eqnarray}
\kappa ^{\mu \nu \rho \sigma }\,F_{\mu \nu }F_{\rho \sigma } &=&\frac{1}{2}\kappa [ g^{\mu \rho }( \xi ^{b}\xi
^{d}e_{\;\;b}^{\nu }e_{\;\;d}^{\sigma }\text{ } -\eta ^{bd}e_{\;\;b}^{\nu
}e_{\;\;d}^{\sigma }\text{ }\,\xi ^{e}\xi _{e}/4) \nonumber \\
&-& g^{\mu \sigma
}( \xi ^{b}\xi ^{c}e_{\;\;b}^{\nu }\,e_{\;\;c}^{\rho }\,-\eta
^{bc}e_{\;\;b}^{\nu }\,e_{\;\;c}^{\rho }\,\,\xi ^{\mu}\xi _{\mu}/4) \text{ 
} \nonumber \\
&+& g^{\nu \sigma }\left( \xi ^{a}\xi ^{c}\,e_{\;\;a}^{\mu
}\,\,e_{\;\;c}^{\rho }\,-\eta ^{ac}\,e_{\;\;a}^{\mu }\,\,e_{\;\;c}^{\rho
}\,\,\xi ^{e}\xi _{e}/4\right) \, \nonumber \\
&-& g^{\nu \rho }( \xi ^{a}\xi
^{d}e_{\;\;a}^{\mu }\,e_{\;\;d}^{\sigma }\text{ }\nonumber \\
&-&\eta ^{ad}e_{\;\;a}^{\mu
}\,e_{\;\;d}^{\sigma }\text{ }\,\xi ^{e}\xi _{e}/4) ] \,F_{\mu\nu }F_{\rho \sigma }. 
\end{eqnarray}
We also have the following relation for tetrad fields 
\begin{eqnarray}
g^{\nu\rho}=\eta^{bc}e^{\nu}_{b}e^{\rho}_{c}=e^{\nu b}e^{\rho}_{c},
\end{eqnarray}
that will lead to the following contribution
\begin{eqnarray}
\kappa ^{\mu \nu \rho \sigma }\,F_{\mu \nu }F_{\rho \sigma } &=&\frac{1}{2}\kappa [ g^{\mu \rho }( \xi ^{\nu }\xi ^{\sigma
} - g^{\nu \sigma }\,\xi ^{e}\xi _{e}/4) \nonumber \\
&-& g^{\mu \sigma }( \xi^{\nu }\xi ^{\rho }\, - g^{\nu \rho }\,\xi ^{e}\xi _{e}/4) \text{ }%
\nonumber \\
&+& g^{\nu \sigma }( \xi ^{\mu }\xi ^{\rho }-g^{\mu \rho }\,\xi ^{e}\xi
_{e}/4) \,\nonumber \\ 
&-& g^{\nu \rho }( \xi ^{\mu }\xi ^{\sigma } \nonumber \\
&-& g^{\mu \sigma
}\,\xi ^{e}\xi _{e}/4) ] \,F_{\mu \nu }F_{\rho \sigma }. 
\end{eqnarray}
We can also rewrite
\begin{eqnarray}
\kappa ^{\mu \nu \rho \sigma }\,F_{\mu \nu }F_{\rho \sigma } &=& \frac{1}{2}\kappa [ \lbrace g^{\mu \rho }( \xi ^{\nu }\xi
^{\sigma }-g^{\nu \sigma }\,\xi ^{e}\xi _{e}/4) \nonumber \\
&+& g^{\nu \sigma }\left(\xi ^{\mu }\xi ^{\rho }-g^{\mu \rho }\,\xi ^{e}\xi _{e}/4\right) \rbrace F_{\mu \nu }F_{\rho \sigma } \nonumber \\
&-&\lbrace g^{\mu \sigma }\left( \xi ^{\nu }\xi^{\rho }\,-g^{\nu \rho }\,\xi ^{e}\xi _{e}/4\right) \text{ }\, \nonumber \\
&+& g^{\nu \rho}\left( \xi ^{\mu }\xi ^{\sigma }-g^{\mu \sigma }\,\xi ^{e}\xi _{e}/4\right)\rbrace \,F_{\mu \nu }F_{\rho \sigma }].\nonumber \\
\end{eqnarray}
Identifying the even symmetry in the terms $\,F_{\mu \nu }F_{\rho \sigma }$ in the pairs $\mu \sigma ,\nu \rho$, and using the antisymmetry in the second term under the exchange $\rho\rightarrow \sigma$, $\sigma\rightarrow \rho$, we achieve
\begin{eqnarray}
\kappa ^{\mu \nu \rho \sigma }\,F_{\mu \nu }F_{\rho \sigma } &=& 2\kappa \left[ g^{\mu \rho }\left( \xi ^{\nu }\xi ^{\sigma }
-g^{\nu\sigma }\,\xi ^{e}\xi _{e}/4\right) F_{\mu \nu }F_{\rho \sigma }\right]. \nonumber \\
\end{eqnarray}
The effective lagrangian now can be written as
\begin{eqnarray}
\mathcal{L}_{\text{modMax}}\ &=&-\sqrt{-g}( \frac{1}{4}\,F_{\mu \nu }F_{\rho \sigma }[
g^{\mu \rho }g^{\nu \sigma } \nonumber \\
&+& 2\kappa g^{\mu \rho }\left( \xi ^{\nu }\xi^{\sigma }-g^{\nu \sigma }\,\xi ^{e}\xi _{e}/4\right) ]). \label{effk2}
\end{eqnarray}
\section{Effective metric}

We can simplify the terms in the lagrangian as follows
\begin{eqnarray}
&& F_{\mu \nu }F_{\rho \sigma }[ g^{\mu \rho }g^{\nu \sigma } \nonumber \\
&& + 2\kappa
g^{\mu \rho }( \xi ^{\nu }\xi ^{\sigma }-g^{\nu \sigma }\,\xi ^{e}\xi
_{e}/4)] \nonumber \\
&& = F_{\mu \nu }F_{\rho \sigma }[ g^{\mu \rho
}g^{\nu \sigma } \nonumber \\
&& + ( 2\kappa g^{\mu \rho }\xi ^{\nu }\xi ^{\sigma
}-\kappa g^{\mu \rho }g^{\nu \sigma }\,\xi ^{e}\xi _{e}/2 )] 
\end{eqnarray}
and rewrite
\begin{eqnarray}
&& F_{\mu \nu }F_{\rho \sigma }\left( g^{\mu \rho }g^{\nu \sigma }+2\kappa
g^{\mu \rho }\left( \xi ^{\nu }\xi ^{\sigma }-g^{\nu \sigma }\,\xi ^{e}\xi
_{e}/4\right) \right) \nonumber \\
&& = F_{\mu \nu }F_{\rho \sigma }\left( g^{\mu \rho }\left( g^{\nu \sigma
}+\kappa \xi ^{\nu }\xi ^{\sigma }\right) +\kappa \xi ^{\mu }\xi ^{\rho
}[ g^{\nu \sigma }-\frac{g^{\nu \sigma }g^{\mu \rho }\,\xi ^{e}\xi
_{e}/2}{\xi ^{\mu }\xi ^{\rho }}\right) ]. \nonumber \\
\end{eqnarray}
The lagrangian can now be written as follows
\begin{eqnarray}
&&\mathcal{L}_{\text{modMax}}\ =-\sqrt{-g}\, \frac{1}{4}\,F_{\mu \nu }F_{\rho \sigma }[ g^{\mu \rho }( g^{\nu \sigma
}+\kappa \xi ^{\nu }\xi ^{\sigma }) \nonumber \\
&& + \kappa \xi ^{\mu }\xi ^{\rho
}( g^{\nu \sigma }-\frac{g^{\nu \sigma }g^{\mu \rho }\,\xi ^{e}\xi
_{e}/2}{\xi ^{\mu }\xi ^{\rho }}) ]. \label{ubluk}
\end{eqnarray}
We can define the effective metric 
\begin{eqnarray}
\tilde{g}_{\mu\nu}=g_{\mu\nu} - \frac{\kappa}{1+\kappa \xi^{\rho}\xi_{\rho}/2}\xi_{\mu}\xi_{\nu}
\end{eqnarray} 
and its inverse as
\begin{eqnarray}
\tilde{g}^{\mu\nu}=g^{\mu\nu} + \frac{\kappa}{1-\kappa \xi^{\rho}\xi_{\rho}/2}\xi^{\mu}\xi^{\nu}.
\end{eqnarray}
In order to prove the identity relation, we have to consider
\begin{eqnarray}
\tilde{g}^{\mu\nu}\tilde{g}_{\mu\nu}&=&\left(g^{\mu\nu} + \frac{\kappa}{1-\kappa \xi^{\rho}\xi_{\rho}/2}\xi^{\mu}\xi^{\nu}\right)\left(g_{\mu\nu} - \frac{\kappa}{1+\kappa \xi^{\rho}\xi_{\rho}/2}\xi_{\mu}\xi_{\nu}\right) \nonumber \\
\end{eqnarray} 
we can simplify
\begin{eqnarray}
\tilde{g}^{\mu\nu}\tilde{g}_{\mu\nu}=g^{\mu\nu}g_{\mu\nu}.
\end{eqnarray} 
The lagrangian can be written after some algebra as follows
\begin{eqnarray}
\mathcal{L}_{\text{modMax}}\ &=&-\sqrt{-g}\, \frac{1}{4}\,F_{\mu \nu }F_{\rho \sigma }\lbrace  \left(1-\kappa \xi^{\mu}\xi_{\mu}/2\right)\tilde{g}^{\mu\rho}\tilde{g}^{\nu\sigma} \nonumber \\
&-& \frac{\kappa^{2}}{1-\kappa \xi^{\mu}\xi_{\mu}/2}\xi ^{\nu }\xi ^{\sigma }\xi^{\mu}\xi^{\rho} \,\rbrace.
\end{eqnarray} 
Taking the small $\kappa$ in the previous result, we can neglect the higher order terms and the lagrangian will reduce to
\begin{eqnarray}
\mathcal{L}_{\text{modMax}}\ &=& -\sqrt{-g}\, \frac{1}{4}\,F_{\mu \nu }F_{\rho \sigma }\left[ \left(1 -\kappa\xi^{\rho}\xi_{\rho}/2 \right)\tilde{g}^{\nu\sigma}\tilde{g}^{\mu\rho} \, \right].
\end{eqnarray}
We could also achieve the approximation for small $\kappa$ by means of the relation
\begin{eqnarray}
&& F_{\mu \nu }F_{\rho \sigma }\left( g^{\mu \rho }\left( g^{\nu \sigma
}+\kappa \xi ^{\nu }\xi ^{\sigma }\right) +\kappa \xi ^{\mu }\xi ^{\rho
}\left( g^{\nu \sigma }-\frac{g^{\nu \sigma }g^{\mu \rho }\,\xi ^{e}\xi
_{e}/2}{\xi ^{\mu }\xi ^{\rho }}\right) \right) \nonumber \\
&\sim &F_{\mu \nu }F_{\rho
\sigma }\left( g^{\mu \rho }\left( g^{\nu \sigma }+\kappa \xi ^{\nu }\xi
^{\sigma }\right) +\kappa \xi ^{\mu }\xi ^{\rho }\left( g^{\nu \sigma
}+\kappa \xi ^{\nu }\xi ^{\sigma }\right) \right),\nonumber \\
\end{eqnarray}
\begin{eqnarray}
\tilde{g}^{\mu \rho } &=&\left( g^{\mu \rho }+\kappa \xi ^{\mu }\xi ^{\rho
}\right) \sim \left( g^{\mu \rho }+\frac{2\kappa }{2+\kappa }\xi ^{\mu }\xi
^{\rho }\right), 
\end{eqnarray}
and then achieving the result
\begin{eqnarray}
\mathcal{L}_{modMax}&=&-\sqrt{-g}\left(1-\frac{1}{2}\kappa\xi^{\rho}\xi_{\rho}\right)\frac{1}{4}F_{\mu\nu}F_{\rho\sigma}\tilde{g}^{\mu\rho}\tilde{g}^{\nu\sigma}. \nonumber \\
\end{eqnarray}	
\section{Momentum Energy Tensor}

The momentum energy tensor from the modified action can be written as
\begin{eqnarray}
{T}_{\alpha \beta }&=& g_{\alpha \beta }\left\{ \frac{1}{4}\left( 1-%
\frac{\kappa }{2}\xi ^{2}\right) F_{\mu \nu }F^{\mu \nu }+\frac{\kappa }{2}%
\xi ^{\nu }\xi ^{\sigma }F^{\rho }\mathrm{\ }_{\nu }F_{\rho \sigma }\right\}
\nonumber \\
&+& F^{\nu }\mathrm{\ }_{\beta }F_{\nu \alpha }\left( 1+\frac{\kappa }{2}\xi
^{2}\right) \nonumber \\
&+& \kappa \xi ^{\nu }\xi ^{\sigma }F_{\alpha \nu }F_{\beta \sigma }.
\end{eqnarray}
The energy associated is then given by
\begin{eqnarray}
{T}_{00} &=& g_{00}\left\{ \frac{1}{4}\left( 1-\frac{\kappa }{2}\xi
^{2}\right) F_{\mu \nu }F^{\mu \nu }+\frac{\kappa }{2}\xi ^{\nu }\xi
^{\sigma }F^{\rho }\ _{\nu }F_{\rho \sigma }\right\} \nonumber \\
&+& F^{\nu }\mathrm{\ }%
_{0}F_{\nu 0}\left( 1+\frac{\kappa }{2}\xi ^{2}\right) +\kappa \xi ^{\nu
}\xi ^{\sigma }F_{0\nu }F_{0\sigma } \nonumber \\
&=&1\left\{ \frac{1}{4}\left( 1-\frac{\kappa }{2}\xi ^{2}\right) \left( 
\mathbf{E}^{2}-\mathbf{B}^{2}\right) +\frac{\kappa }{2}\xi ^{\nu }\xi
^{\sigma }F^{\rho }\mathrm{\ }_{\nu }F_{\rho \sigma }\right\} \nonumber \\
&+& F^{\nu }\
_{0}F_{\nu 0}\left( 1+\frac{\kappa }{2}\xi ^{2}\right) +\kappa \xi ^{\nu
}\xi ^{\sigma }F_{0\nu }F_{0\sigma }.
\end{eqnarray}
\section{Maxwell equations from the modified Maxwell lagrangian}
\begin{figure}[h]
\centering
\includegraphics[scale=0.3]{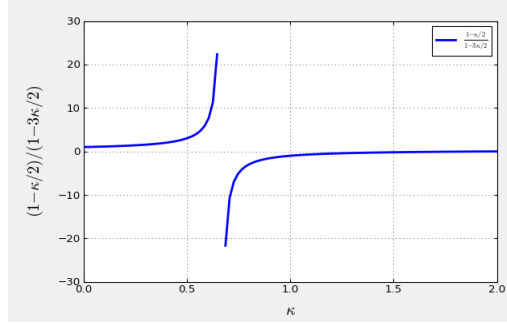}
\caption{(Color online) Term of amplitude dependence on $\kappa$.}
\label{tghiizkl}
\end{figure}
\begin{figure}[h]
\centering
\includegraphics[scale=0.2]{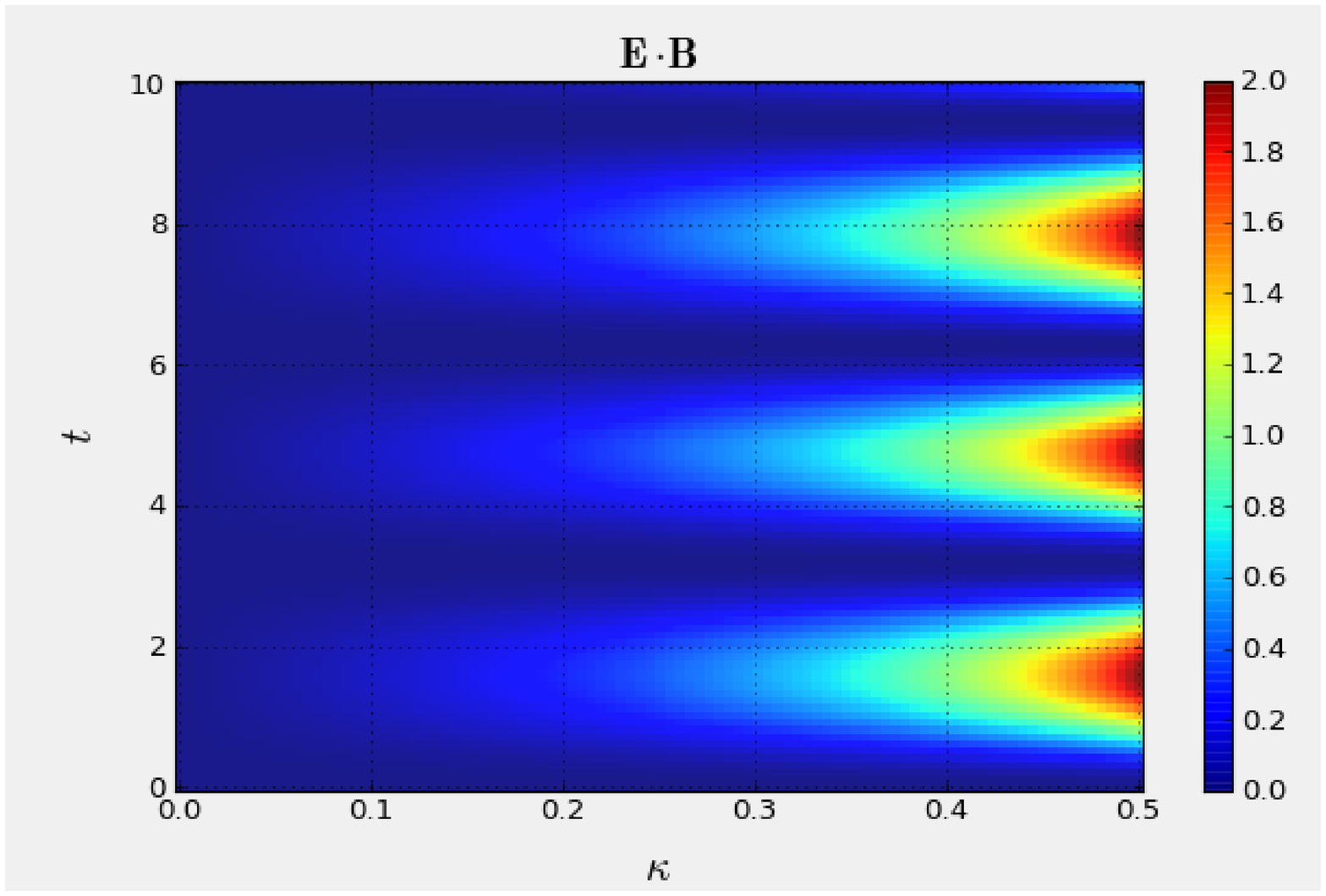}
\includegraphics[scale=0.2]{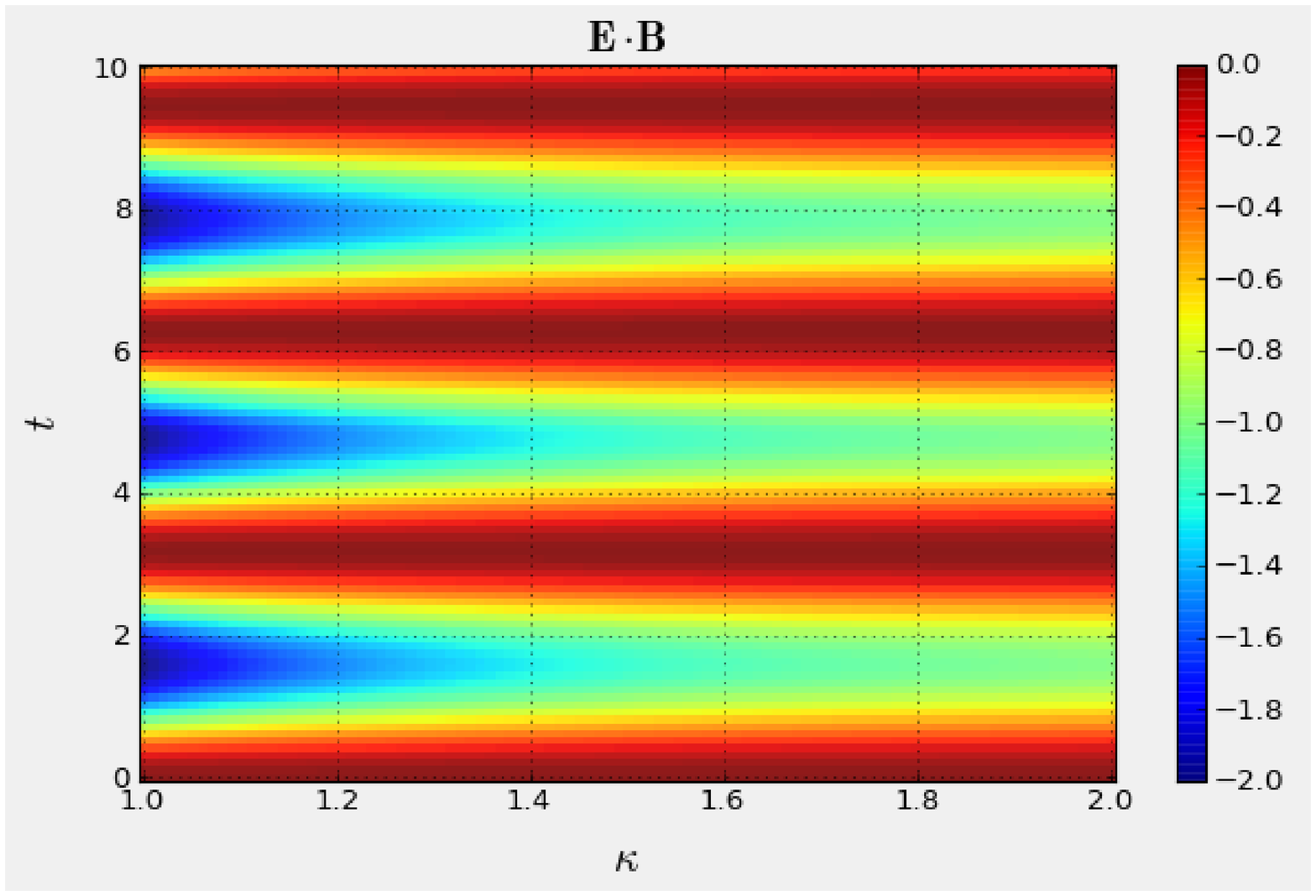}
\includegraphics[scale=0.2]{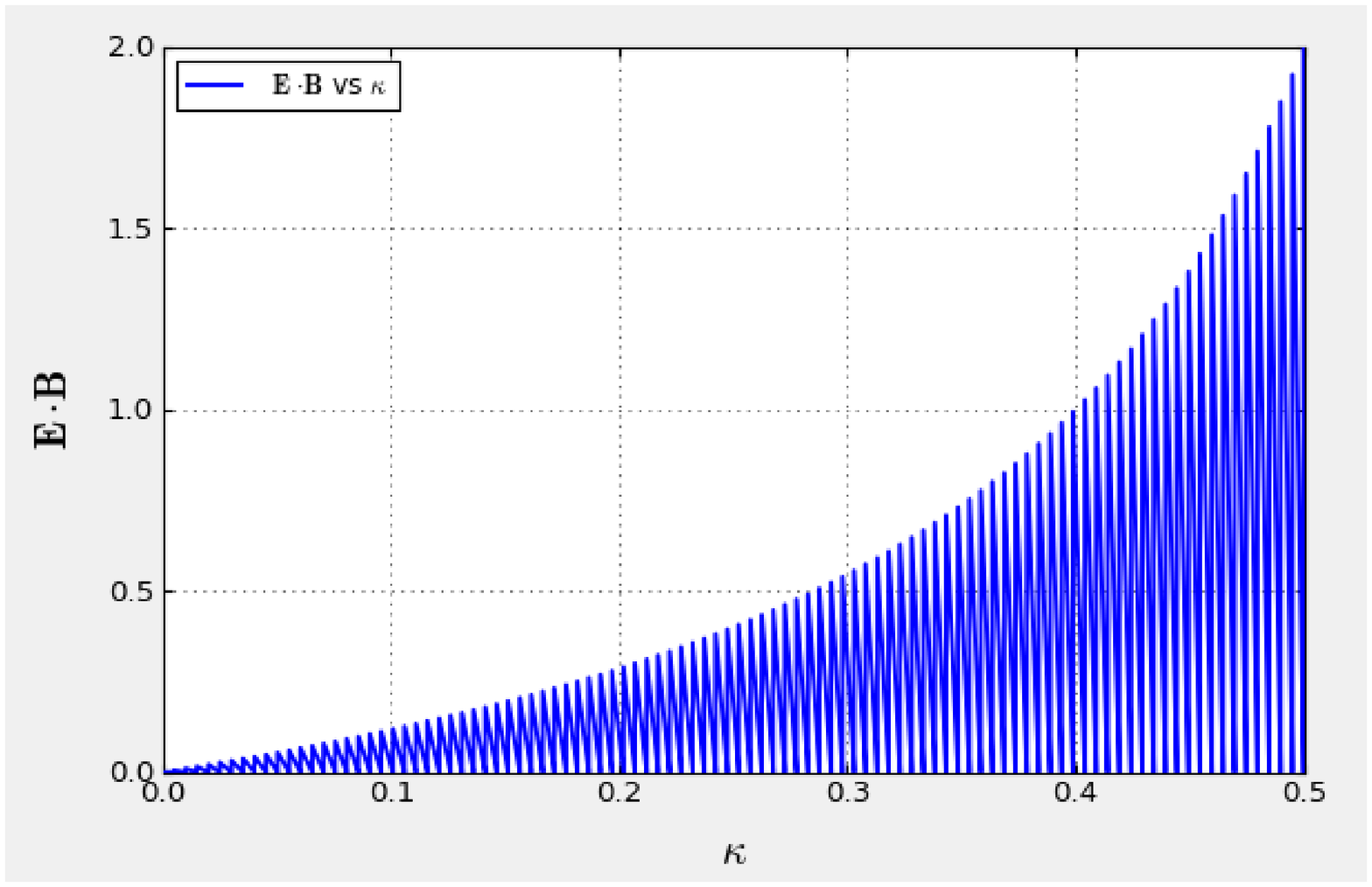}
\includegraphics[scale=0.2]{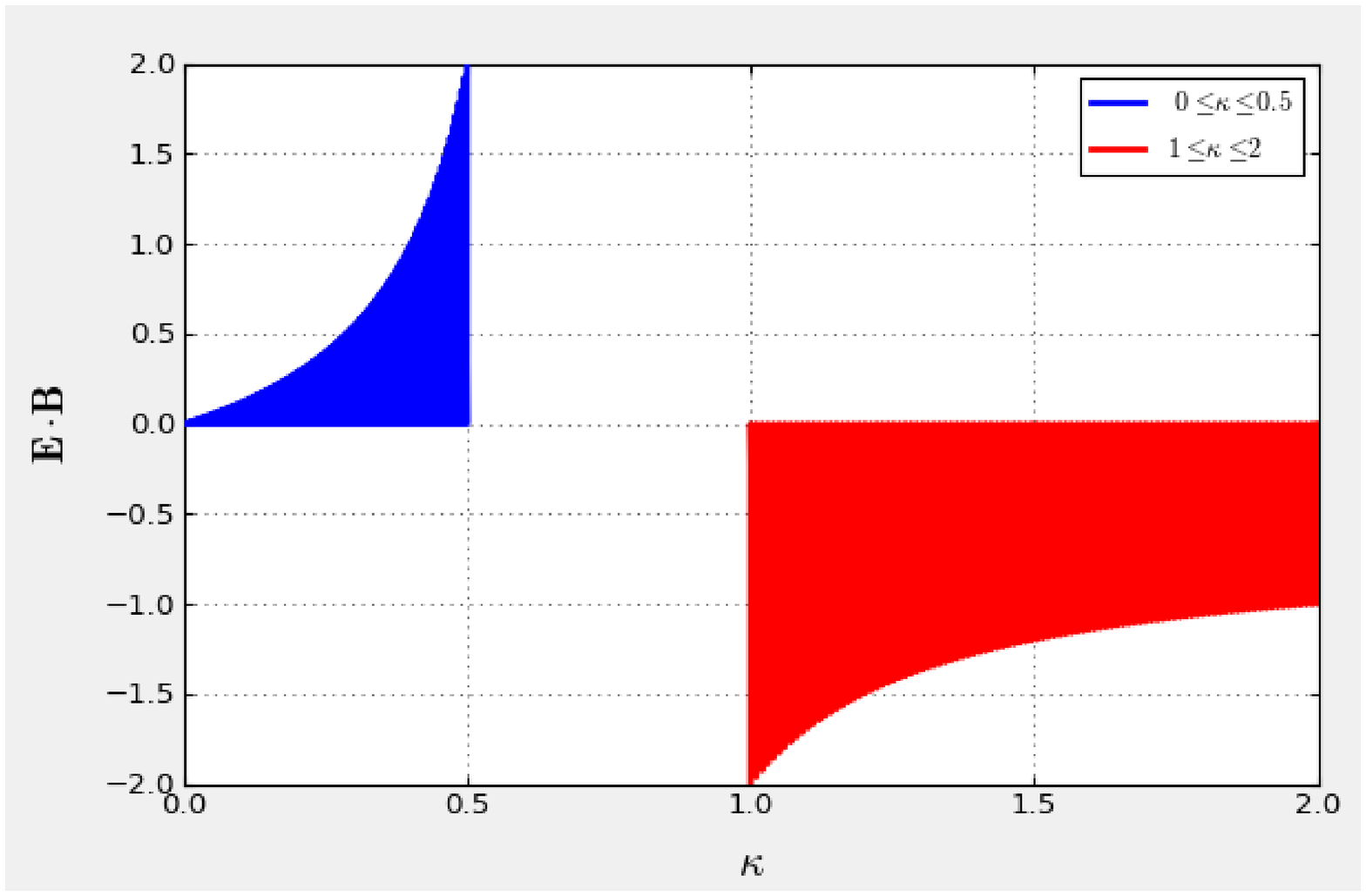}
\caption{(Color online) Behaviour of the scalar product of the linearly polarized electric field with the magnetic fields with 
dependence on $\kappa$, considering $0\leq \kappa \leq 1/2$ and $1\leq \kappa \leq 2$.}
\label{tghiiz91}
\end{figure}
\begin{figure}[h]
\centering
\includegraphics[scale=0.2]{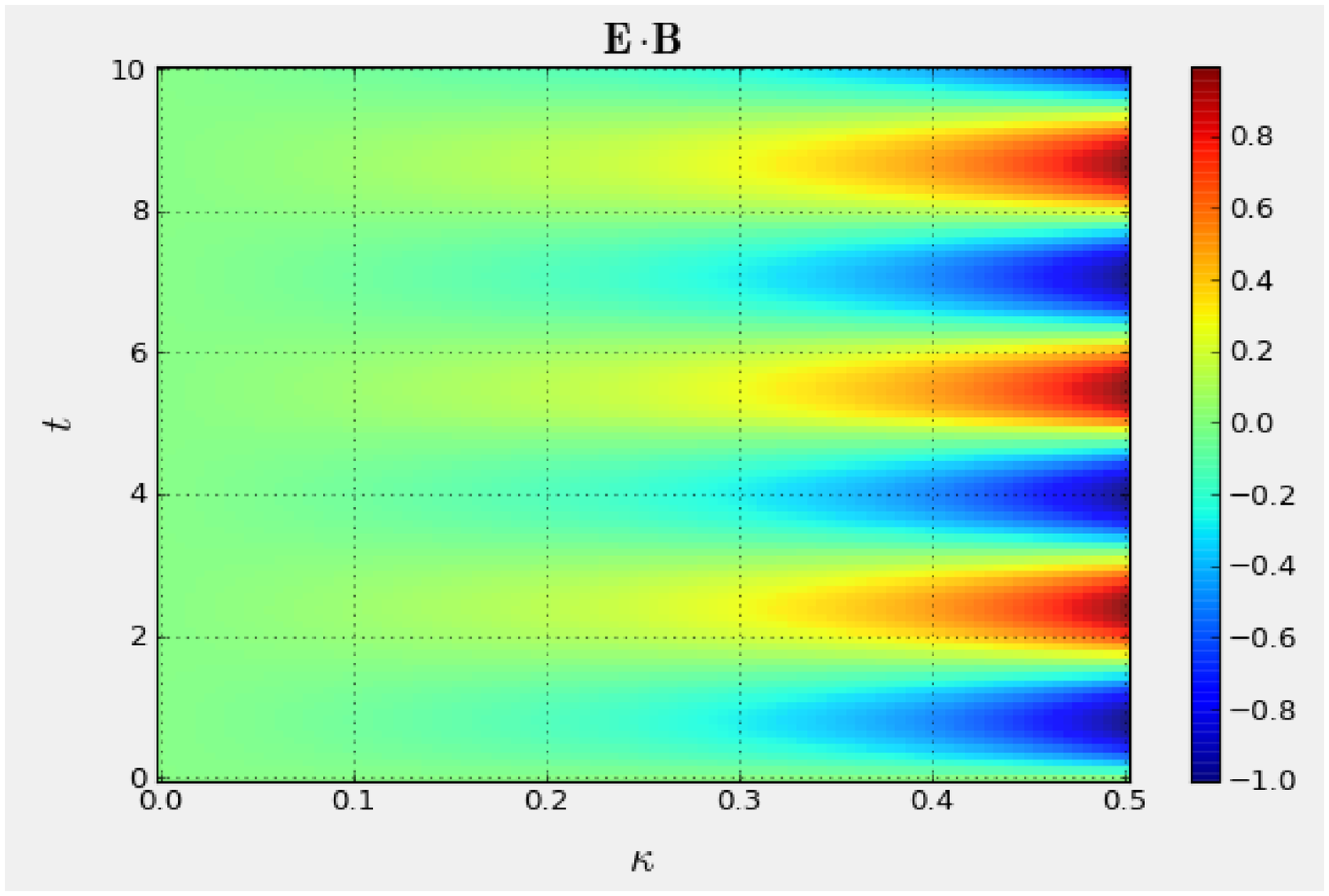}
\includegraphics[scale=0.2]{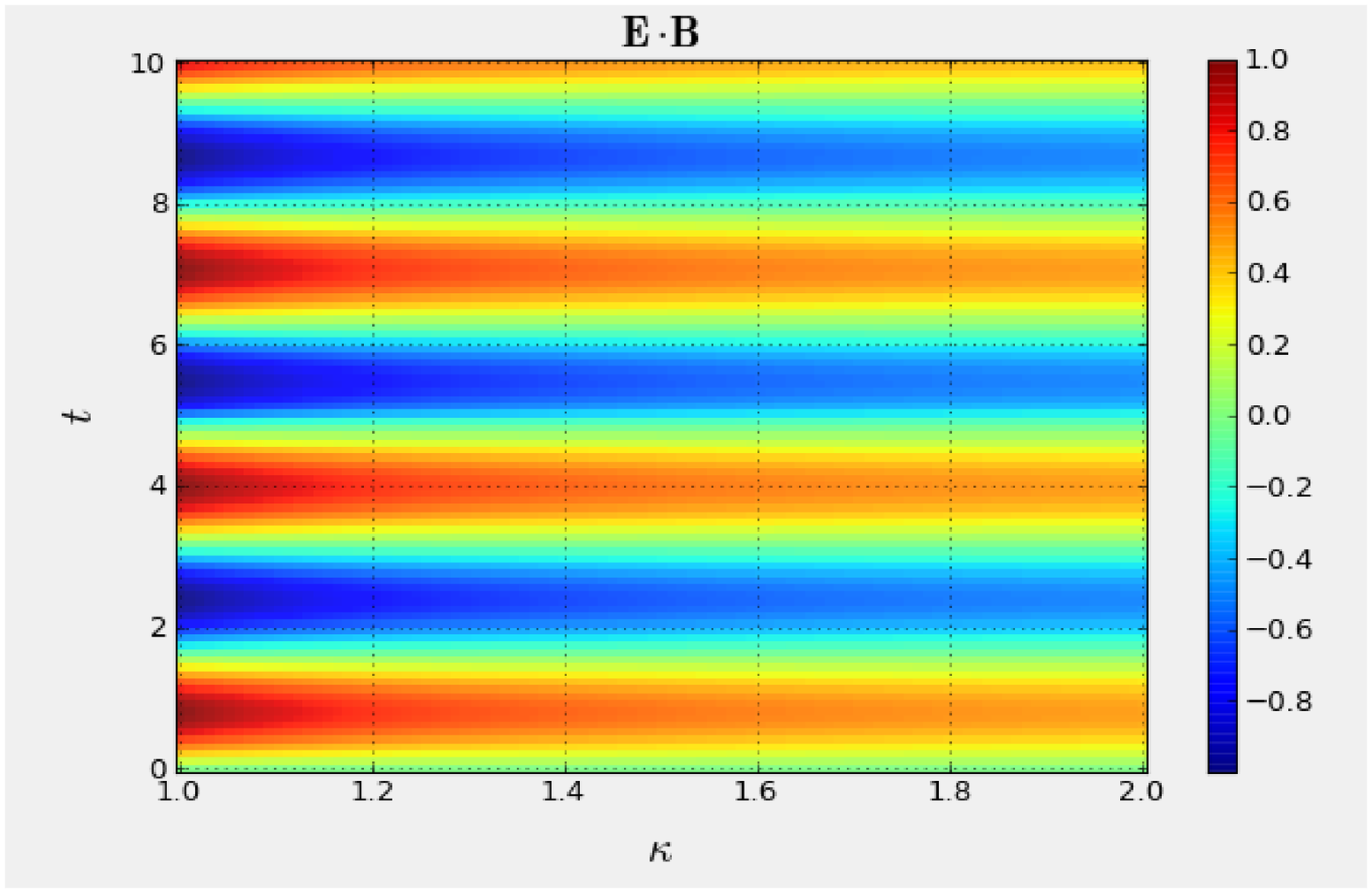}
\includegraphics[scale=0.2]{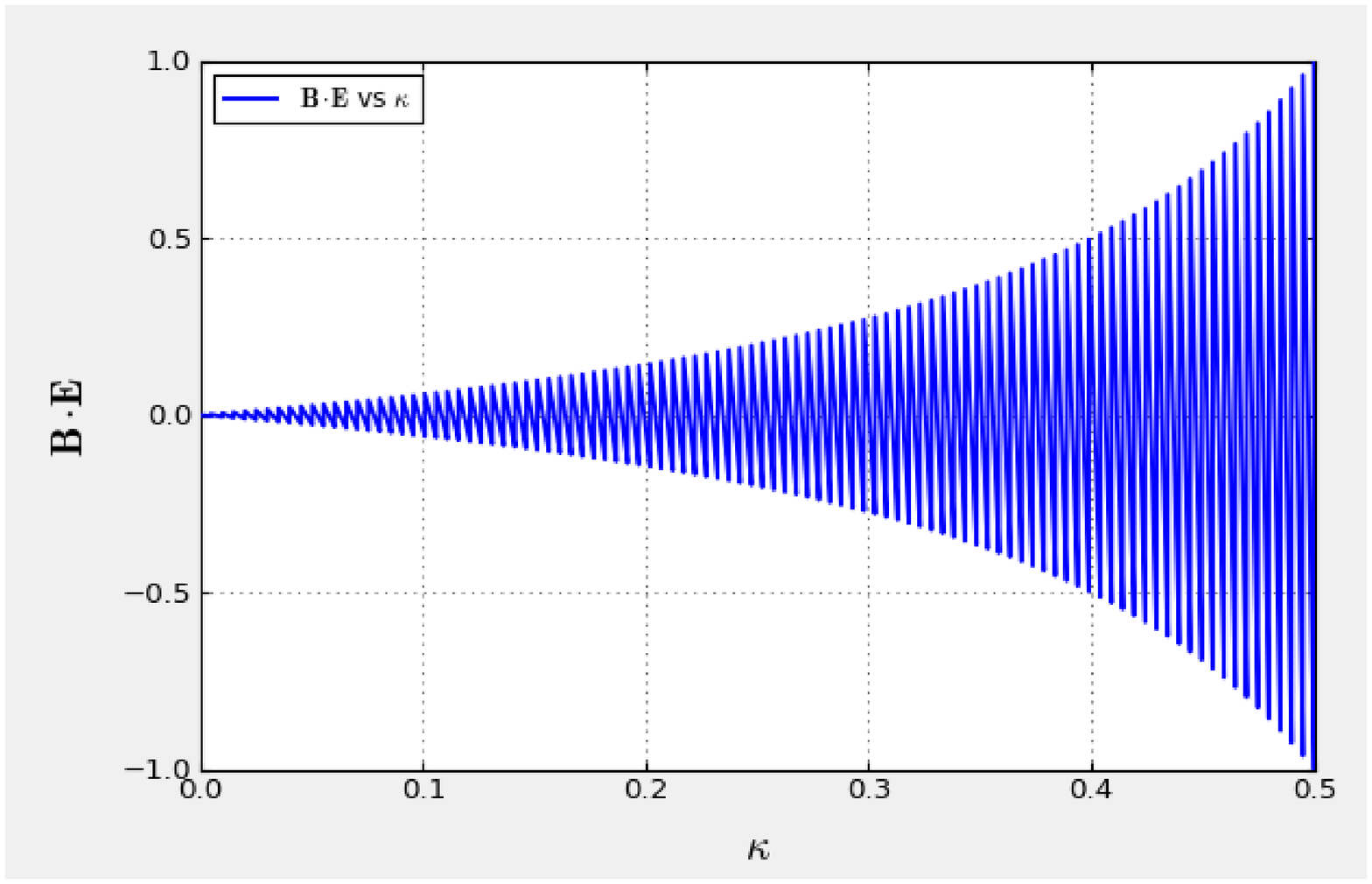}
\includegraphics[scale=0.2]{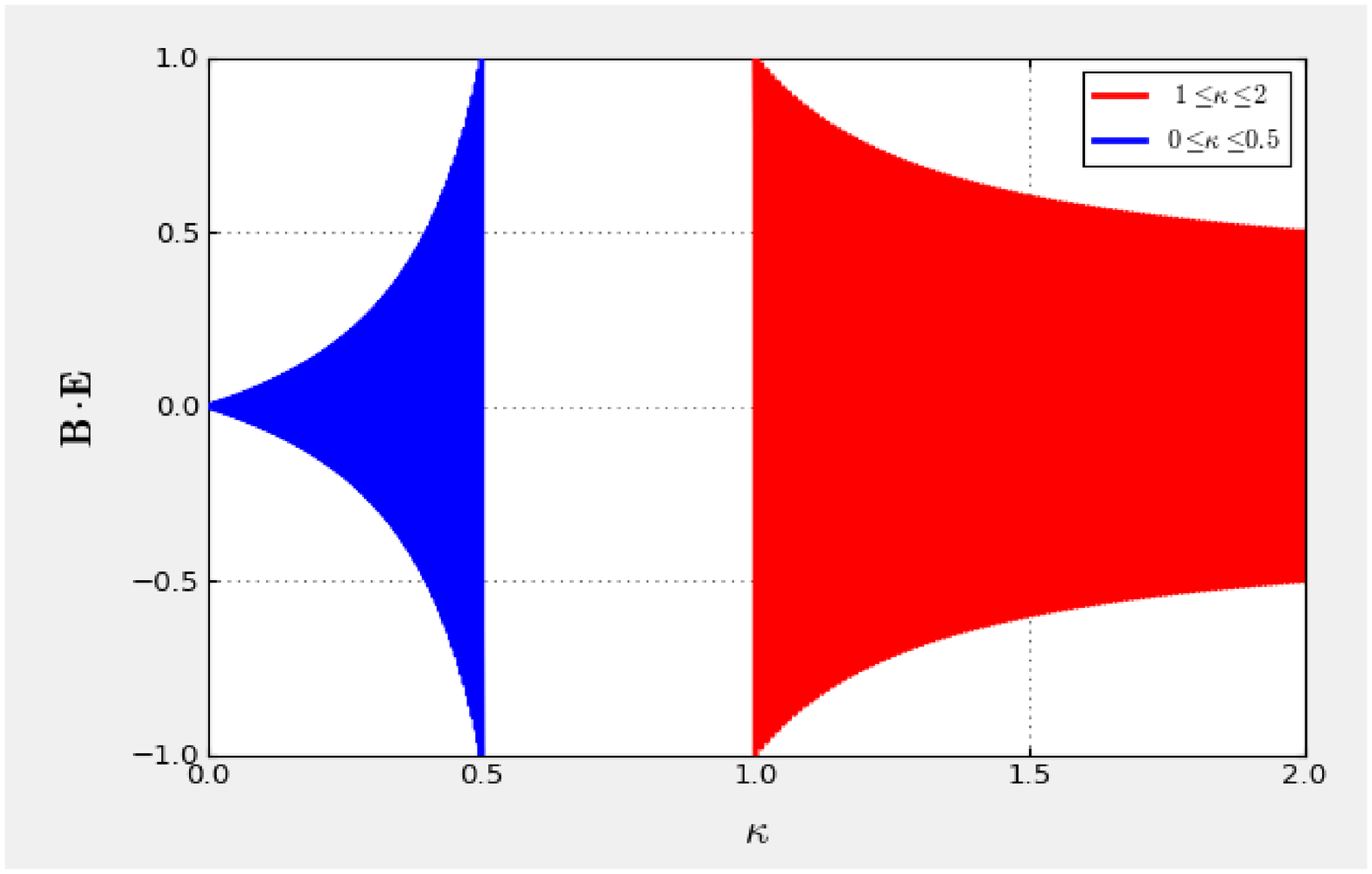}
\caption{(Color online) Scalar product of the circularly polarized electric field with the magnetic fields with 
dependence on $\kappa$, considering $0\leq \kappa \leq 1/2$ and $1\leq \kappa \leq 2$.}
\label{tghiiz99}
\end{figure}

\begin{figure}[h]
\centering
\includegraphics[scale=0.2]{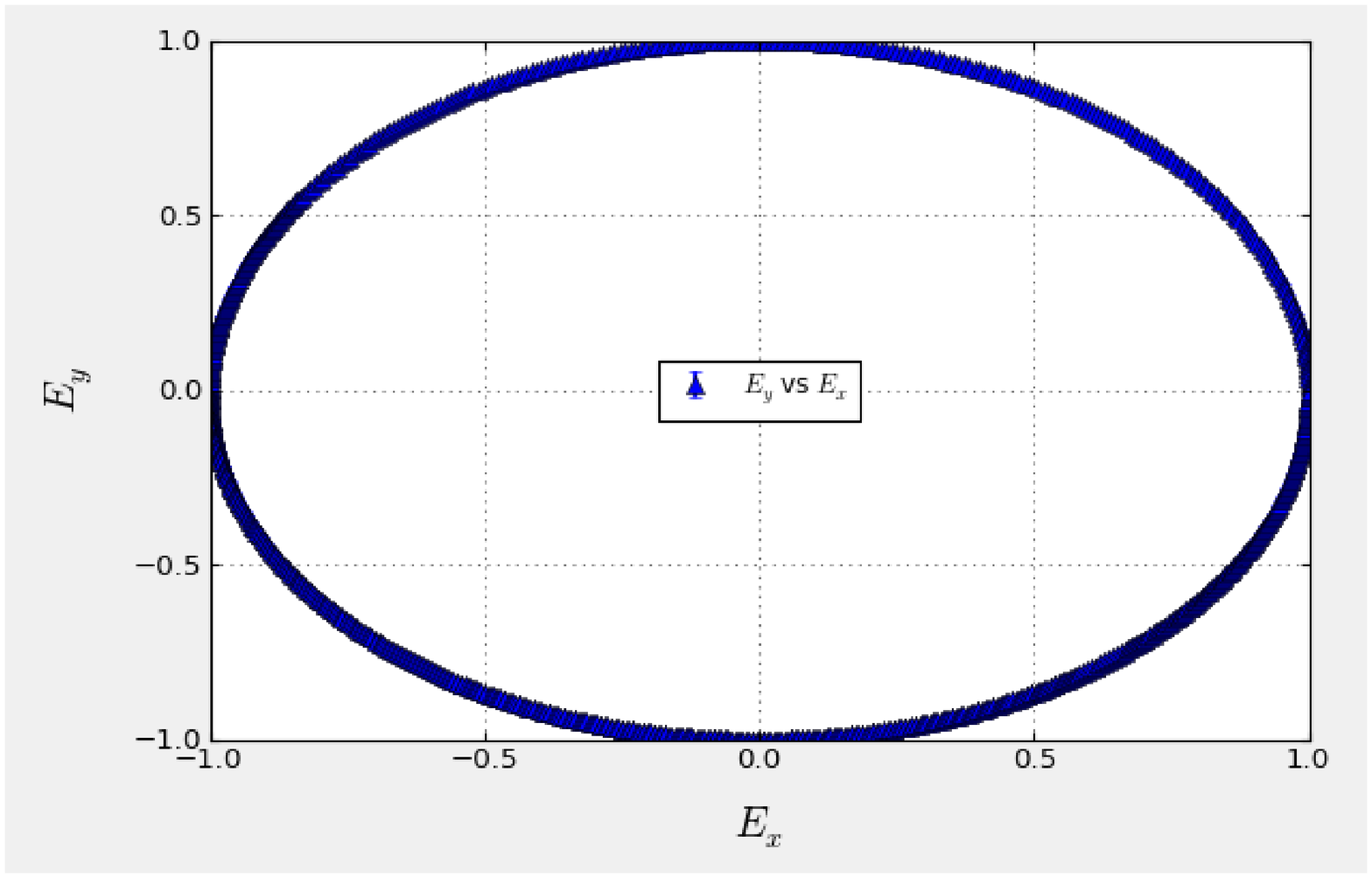}
\includegraphics[scale=0.2]{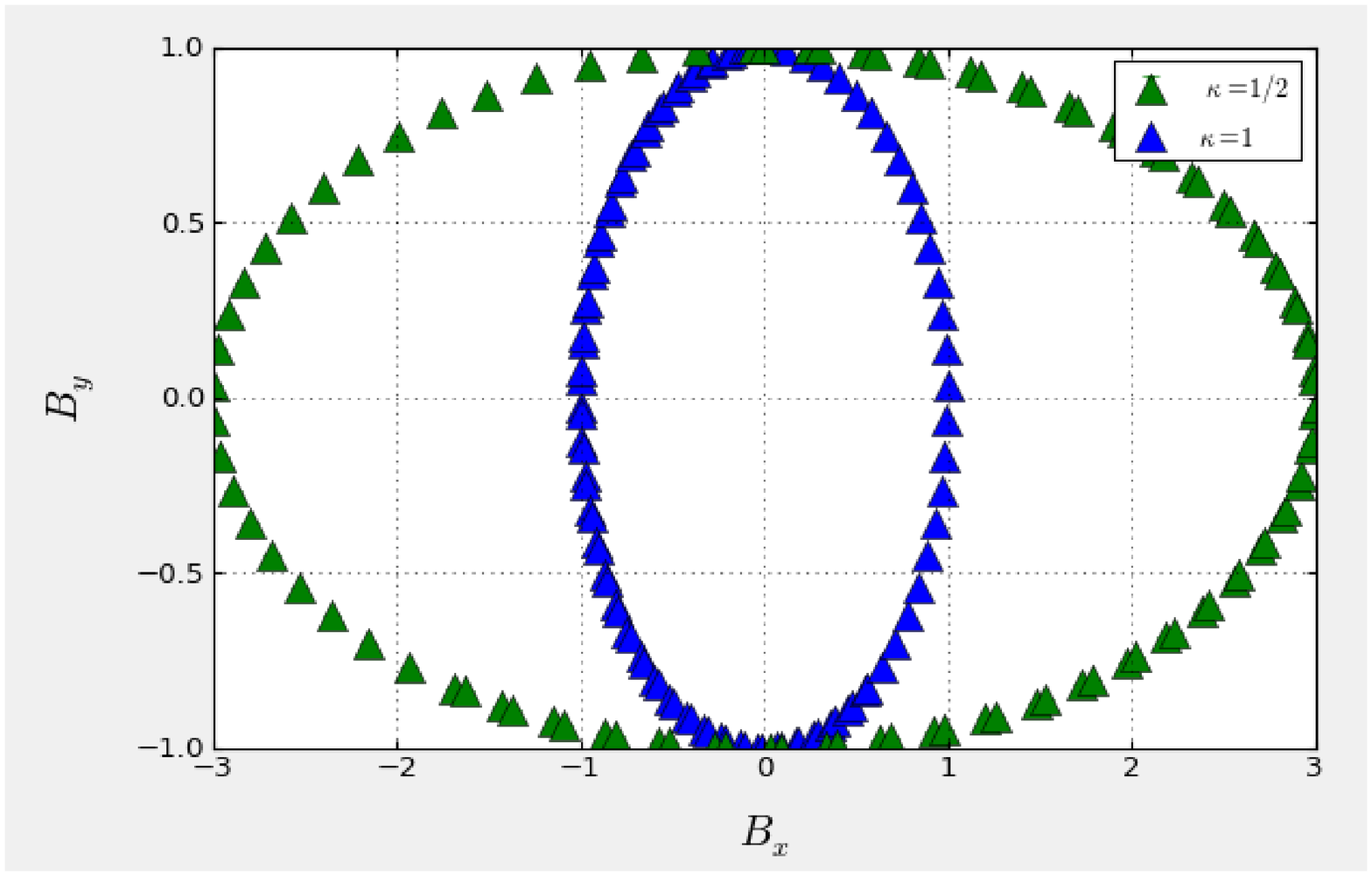}
\caption{(Color online) Effect of the $\kappa$ on the elliptical polarization of the magnetic field, for $\kappa=1$ and 
$\kappa=1/2$.}
\label{tghiiz19}
\end{figure}
The modified Maxwell equations in the vacuum of a $K_{F}$-type Lorentz symmetry breaking theory  in the 
presence of the complex scalar field $\varphi$, and the $\lambda$-parameter associated to a $\lambda|\phi^{4}|$-theory obtained 
in \cite{belich} 
\begin{eqnarray}
-\left(
1-\kappa \,\xi ^{2}/2\right) \left( \partial _{\mu }F^{\mu \nu }\right) &=&\kappa \xi ^{\nu }\xi ^{\sigma }\partial ^{\rho }F_{\rho \sigma }\nonumber \\
&-&\kappa g^{\nu \rho }\xi ^{\mu }\xi ^{\sigma }\partial _{\mu }F_{\rho \sigma
} \nonumber \\
&+& ie\left( \varphi \partial ^{\nu }\varphi ^{\ast }-\varphi ^{\ast }\partial
^{\nu }\varphi \right) \nonumber \\ 
&+& 2e^{2}A^{\nu }\varphi ^{\ast }\varphi.
\label{euu}
\end{eqnarray}%
We write explicitly the modified Maxwell equations, 
\begin{eqnarray}
&-&\left[ \left( 1-\kappa \,\xi ^{2}/2-\kappa \left( \xi ^{0}\right)
^{2}\right) \mathbf{\nabla \cdot }-\kappa \lambda \vec{\xi}\cdot \right] 
\mathbf{E} \nonumber \\
&=&\kappa \xi ^{0}\left( \partial ^{0}\vec{\xi}\cdot \mathbf{E}+
\vec{\xi}\cdot \mathbf{\nabla }\times \mathbf{B}\right) + ie\left( \varphi \partial ^{0}\varphi ^{\ast }-\varphi ^{\ast }\partial
^{0}\varphi \right) \nonumber \\
&+& 2e^{2}\varphi ^{\ast }\varphi \mathbf{\Phi ,} \\
&&\nabla \times E =-\frac{\partial \mathbf{B}}{\partial t},\\
&& \mathbf{\nabla .B}=0,\\
&-&\left( 1-\kappa \,\xi ^{2}/2\right) \left( -\partial _{0}\mathbf{E+\nabla }%
\times \mathbf{B}\right) =\kappa \xi ^{0}\left( \vec{\xi}\mathbf{\nabla .}%
-\kappa \lambda \right) \mathbf{E} \nonumber \\
&-&\kappa \left( \vec{\xi}+\kappa \lambda
\right) \left( \vec{\xi}\cdot \mathbf{\nabla }\times \mathbf{B}\right) \nonumber \\
&-& ie\left( \varphi \mathbf{\nabla }\varphi ^{\ast }-\varphi ^{\ast }\mathbf{%
\nabla }\varphi \right) \nonumber \\
&+& 2e^{2}\mathbf{A}\varphi ^{\ast }\varphi.
\end{eqnarray}
are reduced to the same equation in the absence of the complex scalar field as we start from the modified Maxwell lagrangian in the 
bilinear gauge action we derived
\begin{eqnarray}
\Sigma _{g}&=&\int d^{4}x\lbrace -\frac{1}{4}F_{\mu \nu }F^{\mu \nu } \nonumber \\
&-&\frac{1}{%
4}2\kappa [ ( g^{\mu \rho }( \xi ^{\nu }\xi ^{\sigma } - g^{\nu\sigma }\,\xi ^{e}\xi _{e}/4) ) F_{\mu \nu }F_{\rho \sigma
}] \rbrace ,
\end{eqnarray}%
In this case, we can understand the anisotropy generated by the kind of Lorentz violation
we are considering. We obtain the generalized equation
\begin{eqnarray}
-\partial _{\mu }\left[ (1-\kappa \,\xi ^{e}\xi _{e}/2)F^{\mu \nu }\right]
&=& \partial _{\mu }\left( \kappa F_{\text{ }}^{\mu \sigma }\xi _{\sigma }\xi
^{\nu }\right) \nonumber \\
&-& \partial _{\alpha }\left( \kappa F_{\text{ }}^{\nu \sigma
}\xi _{\sigma }\xi ^{\alpha }\right)
\end{eqnarray}
In terms of the components ($\nu =0$)
\begin{eqnarray}
-\nabla \cdot \left[ (1-\kappa \,\xi ^{e}\xi _{e}/2){\bf E}\right] &=&\partial
_{0}\left( \kappa \xi ^{0}{\bf E}\cdot \vec{\xi}\right) \nonumber \\
&+&\nabla \cdot \left(
\kappa \left( {\bf E}\xi _{0}+{\bf B}\times \vec{\xi}\right) \xi ^{0}\right) \nonumber \\
&-& \partial _{0}\left( \kappa {\bf E}\cdot \vec{\xi}\xi ^{0}\right) \nonumber \\
&-&\nabla
\cdot \left( \kappa \vec{\xi}\left( {\bf E}\cdot \vec{\xi}\right) \right)
\label{38}
\end{eqnarray}
and ($\nu =i$) we have 
\begin{eqnarray}
&&\partial _{0}\left[ (1-\kappa \,\xi ^{e}\xi _{e}/2){\bf E}\right] -\nabla
\times \left[ (1-\kappa \,\xi ^{e}\xi _{e}/2){\bf B}\right] \nonumber \\
&=&\partial_{0}\left( \kappa \vec{\xi}\left( {\bf E}\cdot \vec{\xi}\right) \right) \nonumber \\
&+& \partial _{j}\left( \kappa \left( E_{\text{ }}^{j}\right) \xi _{0}\vec{\xi}%
\right) +\nabla \cdot \left( \kappa \left( \vec{\xi}\times {\bf B}\right) 
\vec{\xi}\right) \nonumber \\ 
&-&\partial _{0}\left( \kappa \left( {\bf E}\right) \xi
_{0}\xi ^{0}+\kappa \left( \vec{\xi}\times {\bf B}\right) \xi ^{0}\right) \nonumber \\
&-&\partial _{j}\left( \kappa \left( {\bf E}\right) \xi _{0}\xi ^{j}+\kappa
\left( \vec{\xi}\times {\bf B}\right) \xi ^{j}\right)
\end{eqnarray}
These equations take into account the possible dynamical variation of the background represented by the presence of $\xi ^{e}$ and $\kappa$. In the static background we achive the usual the modified Maxwell equations in the vacuum of a $K_{F}$-type Lorentz symmetry breaking theory \cite{belich}, in the absence of the complex scalar field $\varphi$, and the $\lambda$-parameter associated to a $\lambda|\phi^{4}|$-theory is zero, is reduced to  modified Maxwell equations by the presence of the $K^{\mu\nu\rho\sigma}$ tensor. In this case, these equations are written
\begin{eqnarray}
\nabla\cdot {\bf E}&=& \frac{-\kappa \xi^{0}}{[1-\kappa \xi^{2}/2 -\kappa (\xi^{0})^{2}]}\vec{\xi}\cdot\frac{\partial {\bf E}}{\partial t} \nonumber \\ 
&+& \frac{-\kappa \xi^{0}}{[1-\kappa \xi^{2}/2 -\kappa (\xi^{0})^{2}]}\vec{\xi}\cdot\nabla\times{\bf B} \label{uk0} \\
\nabla\cdot {\bf B} &=& 0 \label{uk1}\\
\nabla\times {\bf E} &=& -\frac{\partial {\bf B}}{\partial t} \label{uk2} \\
\nabla\times {\bf B} &=& \frac{\partial {\bf E}}{\partial t} + \frac{\kappa \xi^{0}\vec{\xi}\nabla\cdot {\bf E} 
-\kappa \vec{\xi}\left(\vec{\xi}\cdot\nabla \times {\bf B}\right)}{-(1-\kappa \xi^{2}/2)} \label{uk3}
\end{eqnarray}
Notice that these modified Maxwell equations are written in the vacuum by the presence of Lorentz violation in the electromagnetic field dynamics. 
A feature of this modification is that the two equations (\ref{uk1}) and (\ref{uk2}) are not changed.

Considering a $\xi_{\mu}$ point in a specific direction
\begin{eqnarray}
\xi^{0}=0, \vec{\xi}=\hat{\bf x},
\end{eqnarray}
We have $\kappa= 4\tilde{\kappa}^{11}/3$ and
\begin{eqnarray}
\nabla\cdot {\bf E}&=& 0 \label{mju}\\
\nabla\cdot {\bf B} &=& 0 \label{mju2}\\
\nabla\times {\bf E} &=& -\frac{\partial {\bf B}}{\partial t} \label{uk2l} \\
\nabla\times {\bf B} &=& \frac{\partial {\bf E}}{\partial t} +
\frac{\kappa \left(\nabla \times {\bf B}\right)_{x}}{(1-\kappa/2)}\hat{\bf x} \label{uv2l}
\end{eqnarray}
The only change is in the $x$ component of the eq. (\ref{uv2l})
\begin{eqnarray}
(\nabla\times {\bf B})_{x} &=& \frac{\partial E_{x}}{\partial t} +
\frac{\kappa \left(\nabla \times {\bf B}\right)_{x}}{(1-\kappa/2)} \label{uv2lo}
\end{eqnarray}
or
\begin{eqnarray}
(\nabla\times {\bf B})_{x} = \frac{\partial B_{z}}{\partial y} - \frac{\partial B_{y}}{\partial z} &=& \frac{(1-\kappa/2)}{1-3\kappa/2}\frac{\partial E_{x}}{\partial t} \label{ja1} 
\end{eqnarray}

\subsection{Linearly polarized case}

Considering the electromagnetic wave propagating in the $z$ direction, 
we have from the equations (\ref{mju}) and (\ref{mju2})
\begin{eqnarray}
\frac{\partial {E}_{z}}{\partial z}&=& 0, \frac{\partial {B}_{z}}{\partial z}= 0, 
\end{eqnarray}
we can then consider for simplicity $E_{z}=0,B_{z}=0$. Taking the $x$-component of the electric field as monocromatic wave propagation linearly polarized in the $z$ direction
\begin{eqnarray}
E_{x}(z,t)&=& E_{x,0}\sin(kz -\omega t), \\
E_{y}(z,t)&=& E_{y,0}\sin(kz -\omega t).
\end{eqnarray}
We then have from (\ref{ja1})
\begin{eqnarray}
-\frac{\partial B_{y}(z,t)}{\partial z} &=& \frac{(1-\kappa/2)}{1-3\kappa/2}\frac{\partial E_{x}(z,t)}{\partial t} \nonumber \\
&=& -\omega\frac{(1-\kappa/2)}{1-3\kappa/2}E_{x,0}\cos(kz -\omega t).
\end{eqnarray}
The component of the magnetic field will be given by
\begin{eqnarray}
B_{y}(z,t)=B_{y,0}\sin(kz -\omega t),
\end{eqnarray}
from which the relation between electric and magnetic components will be
\begin{eqnarray}
B_{y,0}=\frac{\omega}{k}\frac{(1-\kappa/2)}{1-3\kappa/2}E_{x,0}.
\end{eqnarray}
Notice that there is a deformation factor in the component of the magnetic field due to the presence of Lorentz violation 
(figure \ref{tghiizkl}).
For the other component
\begin{eqnarray}
\frac{\partial B_{x}(z,t)}{\partial z} &=& \frac{\partial E_{y}(z,t)}{\partial t} \nonumber \\
&=& -\omega E_{y,0}\cos(kz -\omega t).
\end{eqnarray}
We have
\begin{eqnarray}
B_{x}(z,t)=B_{x,0}\sin(kz -\omega t),
\end{eqnarray}
then
\begin{eqnarray}
kB_{x,0}\cos(kz -\omega t) &=& \frac{\partial E_{y}(z,t)}{\partial t} \nonumber \\
&=& -\omega E_{y,0}\cos(kz -\omega t).
\end{eqnarray}
and we have
\begin{eqnarray}
B_{x,0} &=& -\frac{\omega}{k} E_{y,0}.
\end{eqnarray}
We then have the magnetic fields
\begin{eqnarray}
B_{x}(z,t)&=& -\frac{\omega}{k} E_{y,0}\sin(kz -\omega t), \\
B_{y}(z,t)&=& \frac{\omega}{k}\frac{(1-\kappa/2)}{1-3\kappa/2}E_{x,0}\sin(kz -\omega t).
\end{eqnarray}
We can calculate the scalar product between the electric and magnetic field
\begin{eqnarray}
{\bf B}\cdot {\bf E}&=& B_{x}(z,t)E_{x}(z,t) + B_{y}(z,t)E_{y}(z,t) \nonumber \\
&=& -\frac{\omega}{k} E_{y,0}E_{x,0}\sin^{2}(kz -\omega t) \nonumber \\
&+& \frac{\omega}{k}\frac{(1-\kappa/2)}{1-3\kappa/2}E_{y,0}E_{x,0}\sin^{2}(kz -\omega t)
\end{eqnarray}
and after simplification
\begin{eqnarray}
{\bf E}(z,t)\cdot {\bf B}(z,t)&=& \frac{\omega}{k}E_{x,0}E_{y,0}\frac{\kappa}{1-3\kappa/2}\sin^{2}(kz -\omega t).\nonumber \\
\end{eqnarray}
This result shows that the effect of Lorentz violation term $\kappa$ is to make the electric and magnetic field non-orthogonals (figure \ref{tghiiz91}). Instead of 
being orthogonals for any time the angle between the eletric and magnetic fields are oscillating on time.

\subsection{Circularly polarized case}

\begin{figure}[h]
\centering
\includegraphics[scale=0.4]{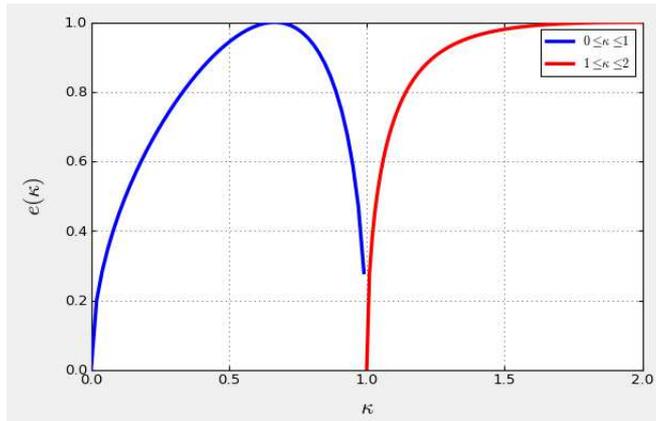}
\caption{(Color online) Eccentricity of the ellipse in elliptical polarization of the magnetic field as a function of $\kappa$, $0 \leq \kappa \leq 2$.}
\label{tghiiz192}
\end{figure}

We can also consider the circularly polarized case, taking the electric field given by
\begin{eqnarray}
E_{x}(z,t)&=& E_{0}\cos(kz -\omega t), \\
E_{y}(z,t)&=& E_{0}\sin(kz -\omega t).
\end{eqnarray}
As before, the component in the direction of propagation is zero. We can also write
\begin{eqnarray}
{\bf E}(z,t)&=& E_{0}\hat{\bf r}(kz -\omega t),
\end{eqnarray}
where
\begin{eqnarray}
\hat{\bf r}(kz -\omega t) &=& \cos(kz -\omega t)\hat{\bf x}+ \sin(kz -\omega t)\hat{\bf y},
\end{eqnarray}
We then have 
\begin{eqnarray}
-\frac{\partial B_{y}(z,t)}{\partial z} &=& \frac{(1-\kappa/2)}{1-3\kappa/2}\frac{\partial E_{x}(z,t)}{\partial t} \nonumber \\
 &=& \omega\frac{(1-\kappa/2)}{1-3\kappa/2}E_{0}\sin(kz -\omega t). \\
\frac{\partial B_{x}(z,t)}{\partial z} &=& \frac{\partial E_{y}(z,t)}{\partial t} \nonumber \\
&=& -\omega E_{0}\cos(kz -\omega t).
\end{eqnarray}
The components of the magnetic field will be
\begin{eqnarray}
B_{x}(z,t)&=& B_{0,x}\sin(kz-\omega t)\\
B_{y}(z,t)&=& B_{0,y}\cos(kz-\omega t)
\end{eqnarray}
and then
\begin{eqnarray}
B_{0,x} &=& \frac{\omega}{k}\frac{(1-\kappa/2)}{1-3\kappa/2}E_{0}, \\
B_{0,y} &=& -\frac{\omega}{k} E_{0}.
\end{eqnarray}
We can then write
\begin{eqnarray}
B_{x}(z,t)&=& \frac{\omega}{k}\frac{(1-\kappa/2)}{1-3\kappa/2}E_{0}\sin(kz-\omega t)\\
B_{y}(z,t)&=& -\frac{\omega}{k} E_{0}\cos(kz-\omega t)
\end{eqnarray}
We then have
\begin{eqnarray}
{\bf E}(z,t)&=& E_{0}\cos(kz -\omega t)\hat{\bf x} \nonumber \\
&+& E_{0}\sin(kz -\omega t)\hat{\bf y}\\
{\bf B}(z,t)&=& \frac{\omega}{k}\frac{(1-\kappa/2)}{1-3\kappa/2}E_{0}\sin(kz-\omega t)\hat{\bf x} \nonumber \\
&-& \frac{\omega}{k} E_{0}\cos(kz-\omega t)\hat{\bf y} 
\end{eqnarray}
In this case we have
\begin{eqnarray}
{\bf E}(z,t)\cdot {\bf B}(z,t) &=& 
\frac{\omega}{k}\frac{(1-\kappa/2)}{1-3\kappa/2}E_{0}^{2}\cos(kz -\omega t)\sin(kz-\omega t) \nonumber \\
&-& \frac{\omega}{k}E_{0}^{2}\cos(kz -\omega t)\sin(kz-\omega t)
\end{eqnarray}
or explicitly
\begin{eqnarray}
{\bf E}(z,t)\cdot {\bf B}(z,t) &=& 
\frac{\kappa}{1-3\kappa/2}\frac{\omega}{k}E_{0}^{2}\cos(kz -\omega t)\sin(kz-\omega t)  \nonumber \\
\end{eqnarray}
We note that the dependence on $\kappa$ makes the electric and magnetic fields non-orthogonal (figure \ref{tghiiz99}) and the elliptical polarization of magnetic field depend on $\kappa$ (figure \ref{tghiiz19}). As a consequence, the eccentricity of the ellipse in the magnetic field is $\kappa$-dependent as depicted in figure \ref{tghiiz192}. The restriction of $\kappa$ to the interval $0 \leq \kappa \leq 2$ respects the limits for unitarity found in \cite{belich}.

\section{Conclusion}

We investigated the influence of CPT-even Lorentz violating contribution of a $K_{F}$-type Lorentz symmetry breaking theory 
tensor in the Maxwell equations derived from the modified Maxwell action of $K_{F}$-Lorentz violation. We considered the features associated to the derivation of the electromagnetic wave solutions in such a Lorentz violation scenario and considered solutions of the cases of linear and elliptical polarized electromagnetic waves propagating in the vacuum. We showed explicitly that the dependence of $\kappa$ of the non-orthogonality between electric and magnetic fields and that the  exihibiting the explicit dependence on $\kappa$. 

We showed in this case the Lorentz violation changes the amplitude of the magnetic field with respect on $\kappa$, without changing the 
electric field. As a consequence, we showed that the electromagnetic wave fields have a non-orthogonal propagation in the vacuum and exhibited an explicity dependency of the eccentricity of the magnetic field polarization on 
$\kappa$-term. Furthermore, we exhibited numerically the consequences of this effect in the cases of linear and elliptical
polarization, in particular, the regimes of non-orthogonality of the electromagnetic wave fields and the eccentricity of the elliptical polarization of the magnetic field with dependence on the $\kappa$-term.
 
\section{Acknowledgements}

The authors acknowledge the support by projects FAPEMA-PRONEM-01852/14, FAPEMA-APCINTER-00273/14, Enxoval–UFMA PPPG N03/2014 (Brazil), UFMA-Res.No1342-CONSEPE Art1-III-1150/2015-33, UFMA-Res.No1342-CONSEPE Art1-IV-1151/2015-88, CNPq (Brazil) and and FAPEMA-UNIVERSAL-01401/16. \\ T. P. also thanks Federal University of Espirito Santo (UFES) \\ for the hospitality in his visit and where this work were concluded as part of the colaboration FAPEMA PRONEM-01852/14.

\end{document}